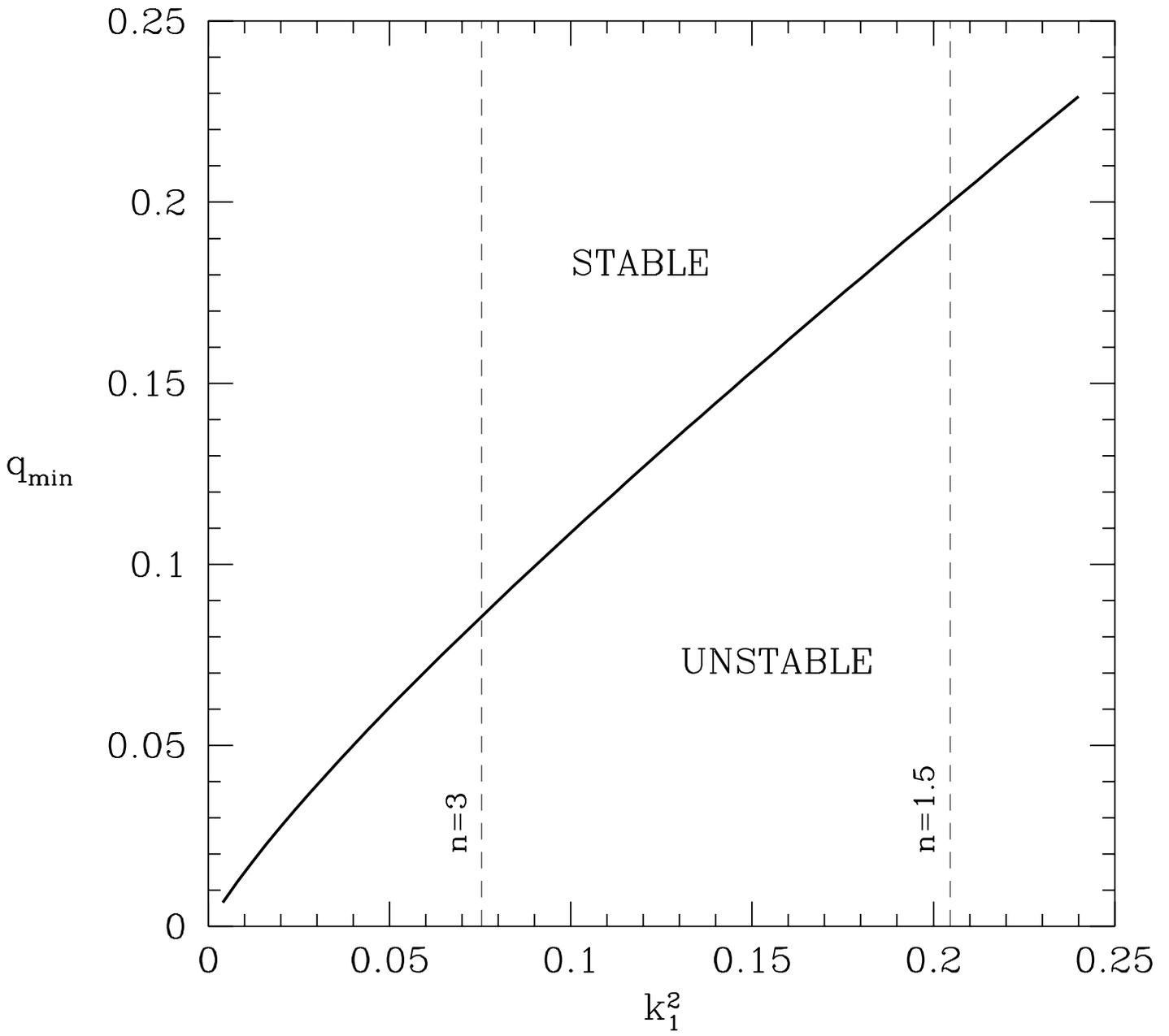

# The Minimum Mass Ratio of W Ursae Majoris Binaries

Frederic A. Rasio[1]

Institute for Advanced Study, Olden Lane, Princeton, NJ 08540

Email: rasio@guinness.ias.edu

## ABSTRACT

The minimum mass ratio for tidal stability of a contact binary containing two unevolved main-sequence stars is calculated to be $q_{\rm min} \simeq 0.09$ in the case of a mostly radiative primary, and is higher if an appreciable fraction of the mass lies in a convective envelope. At least one observed system, AW UMa, has a mass ratio just below this value ($q = 0.075$), implying, if the system is stable, that the primary must be slightly evolved and must have a very shallow convective envelope. Contact binaries with mass ratios significantly below that of AW UMa should not be observed, since they are tidally unstable and quickly merge into a single, rapidly rotating object, on a timescale $\sim 10^3$–$10^4$ yr.

*Subject headings:* hydrodynamics – instabilities – stars: rotation – stars: binaries: close – stars: blue stragglers – stars: evolution

## 1. Introduction

It is estimated that perhaps as much as 1 in every 150 stars in the Galaxy is a contact binary of the W UMa type, containing two main-sequence stars embedded in a thin common envelope of gas (Rucinski 1994). The structure and evolution of these systems remains a major unsolved problem in stellar astrophysics. After heated debates between the proponents of various theoretical models in the late seventies (e.g., Shu 1980 for a brief account) the field went dormant, with a number of fundamental theoretical issues remaining unresolved.

On the observational side, however, contact binaries continue to be discovered in large numbers and the quality of the data is improving steadily. It is likely that the rate of new detections will continue to increase in the near future. In particular, an important byproduct of the recent searches for gravitational microlensing by MACHOS will be a large catalog of eclipsing binaries detected under very uniform conditions (Udalski et al. 1994).

---

[1] Hubble Fellow.

Interest in W UMa binaries was revived recently with the discovery of large numbers of new ones among blue stragglers in open and globular clusters (Kaluzny & Shara 1988; Kaluzny 1990; Mateo et al. 1991; Yan & Mateo 1994). It appears likely that at least some of the blue stragglers are formed by the merging of contact binaries. From a theoretical point of view, this is an important example of the basic hydrodynamic process of binary coalescence, which is relevant to a wide variety of topical problems, including Type Ia supernovae, gamma-ray bursts, and the detection of gravitational waves (Rasio & Shapiro 1994, 1995).

In this *Letter*, I discuss a particular mechanism that can drive the coalescence of W UMa binaries: secular tidal instablilities. I examine the consequences of these instabilitities for the observed distribution of binary mass ratios, showing that the existence of a minimum mass ratio $q_{\rm min} \lesssim 0.1$ can be explained very naturally.

## 2. Minimum Mass Ratio for Stability

The combination of tidal interaction and viscous dissipation in a close binary normally drives the system toward a synchronous (uniformly rotating) configuration. However, if the amount of angular momentum in the orbital motion is below a certain critical value, the stellar spins can no longer remain synchronized and an instability develops. This instability is secular, growing on the viscous dissipation timescale. It leads to orbital decay and, ultimately, merging of the two stars into a single, rapidly rotating object.

Secular tidal instabilities in close binary systems, discovered over a century ago by Darwin (1879), have been discussed more recently by a number of authors (Kopal 1972; Counselman 1973; Hut 1980). Pringle (1974) and Levine et al. (1993) point out their importance for understanding the orbital decay of massive X-ray binaries. Lai, Rasio, & Shapiro (1994) give a very general treatment for close binaries containing polytropic components. The stability condition in general is that along a sequence of synchronized binary configurations with decreasing separation the onset of instability corresponds to the point where the total angular momentum is minimum. In the particular case of a system containing two rigid spheres, it is trivial to show that this is equivalent to the more familiar criterion (Hut 1980) that the orbital angular momentum must exceed three times the total spin angular momentum for stability.

Consider a contact binary with small mass ratio $q = M_2/M_1 \ll 1$. We can then neglect the spin angular momentum of the small, compact secondary. Let's assume that the mass distribution in the primary is centrally condensed, so that its moment of inertia is close to that of the unperturbed (spherical) star. In practice this is a good approximation if only a small fraction of $M_1$ is in a convective envelope (see below). We can then write the orbital angular momentum $J_{\rm orb}$ and the spin angular momentum $J_{\rm spin}$ as

$$J_{\rm orb} = \mu a^2 \Omega; \qquad J_{\rm spin} = M_1 k_1^2 R_1^2 \Omega. \qquad (1)$$

Here $a$ is the binary separation, $\Omega$ is the orbital angular frequency, $k_1$ is the dimensionless gyration radius of the primary, and $\mu = M_1 q/(1+q)$ is the reduced mass. The onset of instability corresponds to $J_{\rm orb} = 3 J_{\rm spin}$, or a critical separation $a_{\rm inst}$ given by

$$\frac{a_{\rm inst}}{R_1} = \left[\frac{3(1+q)}{q}\right]^{1/2} k_1. \qquad (2)$$

This must be compared to the separation $a_{\rm RL}$ at the Roche limit, for which we can write

$$\frac{a_{\rm RL}}{R_1} = \frac{a_{\rm RL}}{\bar{R}_{\rm L1}} = \frac{0.6 + q^{2/3} \ln(1 + q^{-1/3})}{0.49}, \qquad (3)$$

using Eggleton's (1983) fitting formula for the volume mean radius $\bar{R}_{\rm L1}$ of the primary's Roche lobe.

Setting $a_{\rm inst} = a_{\rm RL}$ and solving (numerically) for $q$, we obtain the *minimum mass ratio for stability*, $q_{\rm min}$. A system with $q < q_{\rm min}$ cannot exist in a stable contact configuration. The results are plotted in Figure 1 as a function of $k_1$, the only parameter in this simple model. To within $\sim 20\%$, one has the remarkably simple approximate solution $q_{\rm min} \simeq k_1^2$.

The results obtained from equations (1)–(3) are only valid as long as $q_{\rm min} \ll 1$, and for a centrally condensed primary. In general, fully three-dimensional numerical calculations must be performed. This was done recently by Rasio & Shapiro (1995) for a system containing two fully convective MS stars (modelled as $n = 1.5$ polytropes with the polytropic constants adjusted so as to obtain a simple mass-radius relation with $R \propto M$). For such a system it is found that $q_{\rm min} \simeq 0.45$. As can be seen in Figure 1, the rigid sphere and Roche approximations fail badly in this case, underestimating $q_{\rm min}$ by about a factor 2. For binaries containing stars even less compressible than $n = 1.5$ polytropes, such as neutron stars, it is possible to find that the binary always becomes secularly unstable before reaching contact, for all mass ratios (Rasio & Shapiro 1994). However, for sufficiently compressible stars, such as a mostly radiative main-sequence star (well modeled by a polytrope with $n = 3$, $\Gamma_1 = 5/3$) numerical results indicate that equations (1)–(3) are an excellent approximation (Rasio 1995).

## 3. Discussion

Among the $\sim 100$ contact binaries with reliable photometric data, the system AW UMa has the smallest observed mass ratio, $q = 0.075$ (Rucinski 1992b). Recently van't Veer (1994) pointed out that the absence of observed systems with mass ratios below that of AW UMa cannot be a selection effect, and is therefore unexplained. Indeed, in the literature on contact binaries, it has always been assumed that the minimum mass ratio allowed for stability was much smaller, $q_{\min} \simeq 0.002$ (see, e.g., Rucinski 1985, §3.1.5). This number can be traced back to the pioneering work by Webbink (1976, §5). Unfortunately, the criterion used by Webbink (1976), that the *primary radius* be maximum at the onset of instability, is not the correct one. For a star with $k_1^2 = 0.1$, the value adopted by Webbink (1976), the correct criterion (derived in §2) gives $q_{\min} \simeq 0.11$.

To obtain $q_{\min} = 0.075$ (placing AW UMa just at the stability boundary), the primary must have $k_1^2 \simeq 0.06$. This is less than the value for an $n = 3$ polytrope, which has $k^2(n=3) = 0.08$. The immediate implication is that the primary in AW UMa cannot have much of a convective envelope, and must be slightly evolved. The latter point is consistent with this system being classified observationally as an "A-type" system (primary eclipse is a transit of the secondary; Rucinski 1985). The former is in conflict with at least some simple models for the interior structure of contact binaries (Rucinski 1992a). Note that a convective envelope is not required for the existence of a contact configuration: contact binaries containing O and B stars have been observed (Figueiredo, De Greve, & Hilditch 1994).

The broader implication, of course, is that systems with mass ratios below that of AW UMa may not be observed simply because they are unstable and undergo rapid merging. The timescale for merging can be estimated from standard tidal dissipation theory. The orbital decay time $t_D$ for an unstable system is comparable to what would be the circularization time if it were stable. Using the expressions given by Zahn (1992) we get $t_D \sim 10^3 - 10^4$ yr (the lower number corresponding to eddy viscosity in a convective envelope, the higher number to radiative damping of dynamical tides). This is much shorter than the typical timescale $t_O$ of observed orbital decay in contact binaries. For example, in AW UMa, cumulative changes in the times of eclipses (the "O−C" curve), when interpreted as a continuous orbital period evolution, imply that the orbit is decaying on a timescale $t_O = P_O/(-\dot{P}_O) \simeq 3 \times 10^6$ yr (Demircan, Derman, & Müyesseroğlu 1992). This is much longer than $t_D$, indicating that the observed period changes are almost certainly not related to tidal instability. The inequality $t_O/t_D \gg 1$ in general would imply that the predicted edge in the distribution of observed mass ratios at $q = q_{\min}$ should be very sharp.

- 5 -

In general the stability of a contact binary with given masses $M_1$ and $M_2$ depends on its exact interior structure, and on the degree of contact, or filling factor, $0 < F < 1$ ($F$ is defined in terms of the value of the effective potential at the fluid surface; see Rucinski 1985, §3.1.6). Systems in deeper contact are more susceptible to instabilities, suggesting that the preference for very shallow contact among observed systems (most have $0 < F < 0.25$) may also be related to instabilities (Rasio 1994). For example, Rasio & Shapiro (1995) find that a contact system made of two identical $n = 1.5$ polytropes is tidally unstable for $F \gtrsim 0.2$. Detailed comparisons between the stability boundaries in the parameter space $(M_1, M_2, F)$, predicted theoretically for various interior models, and the distribution of observed systems could provide very tight constraints on the theoretical models.

I am very grateful to S. Ruciński for many useful discussions. I also thank P. Eggleton, B. Pacyński, F. van't Veer, and R. Wijers for comments. This work has been supported by a Hubble Fellowship, funded by NASA through Grant HF-1037.01-92A from the Space Telescope Science Institute, which is operated by AURA, Inc., for NASA, under contract NAS5-26555.

## REFERENCES

Counselman, C. C. 1973, ApJ, 180, 307

Darwin, G. H. 1879, Phil. Trans. R. Soc., 170, 1

Demircan, O., Derman, E., & Müyesseroğlu, Z. 1992, A&A, 263, 165

Eggleton, P. P. 1983, ApJ, 268, 368

Figueiredo, J., De Greve, J. P., & Hilditch, R. W. 1994, A&A, 283, 144

Hut, P. 1980, A&A, 92, 167

Kaluzny, J. 1990, Acta Astron., 40, 61

Kaluzny, J., & Shara, M. M. 1988, AJ, 95, 785

Kopal, Z. 1972, ApSS, 17, 161

Lai, D., Rasio, F. A., & Shapiro, S. L. 1994, ApJ, 423, 344

Levine, A., Rappaport, S., Deeter, J. E., Boynton, P. E., & Nagase, F. 1993, ApJ, 410, 328

Mateo, M., Harris, H. C., Nemec, J., & Olszewski, E. W. 1990, AJ, 100, 469

Pringle, J. E. 1974, MNRAS, 168, 13P

– 6 –Rasio, F. A. 1994, in Evolutionary Links in the Zoo of Interactive Binaries, ed. F. D'Antona et al. (Mem. S. A. It.), 37

Rasio, F. A. 1995, in preparation

Rasio, F. A., & Shapiro, S. L. 1994, ApJ, 432, 242

Rasio, F. A., & Shapiro, S. L. 1995, ApJ, 438, 887

Rucinski, S. M. 1985, in Interacting Binary Stars, ed. J. E. Pringle & R. A. Wade (Cambridge: Cambridge Univ. Press), 85

Rucinski, S. M. 1992a, AJ, 103, 960

Rucinski, S. M. 1992b, AJ, 104, 1968

Rucinski, S. M. 1994, PASP, 106, 462

Shu, F. H. 1980, in Close Binary Stars: Observations and Interpretation, IAU Symp. 88, ed. M. J. Plavec, D. M. Popper, & R. K. Ulrich (Dordrecht: Reidel), 477

Udalski, A., Kubiak, M., Szymański, M., Kaluzny, J., Mateo, M., & Krzemiński, W. 1994, Acta Astron., 44, 317

van't Veer, F. 1994, in Evolutionary Links in the Zoo of Interactive Binaries, ed. F. D'Antona et al. (Mem. S. A. It.), 105

Webbink, R. F. 1976, ApJ, 209, 829

Yan, L., & Mateo, M. 1994, AJ, 108, 1810

Zahn, J.-P. 1992, in Binaries as Tracers of Stellar Formation, ed. A. Duquennoy & M. Mayor (Cambridge: Cambridge Univ. Press), 253This preprint was prepared with the AAS LaTeX macros v3.0.

Fig. 1.— The minimum mass ratio for stability $q_{\min}$ calculated for a spherical primary in the Roche approximation (solid line). The parameter $k_1^2$ is the dimensionless gyration radius of the primary (values for $n = 1.5$ and $n = 3$ polytropes are indicated by the vertical dashed lines).